\documentclass[apb,prl,twocolumn,showpacs]{revtex4-1}
\usepackage{graphicx}
\usepackage{epsfig}
\usepackage{amsmath}

\begin{document}

\title{Solvent induced  current-voltage hysteresis and negative differential resistance in molecular junctions}

\author{Alan A. Dzhioev}
\altaffiliation{On leave of absence from Bogoliubov Laboratory of Theoretical Physics, Joint Institute for Nuclear Research,  RU-141980 Dubna, Russia }
\affiliation{Department of Physics, Universit\'e Libre de Bruxelles, Campus Plaine, CP 231, Blvd du Triomphe, B-1050 Brussels, Belgium }

\author{D. S. Kosov}
\affiliation{Department of Physics, Universit\'e Libre de Bruxelles, Campus Plaine, CP 231, Blvd du Triomphe, B-1050 Brussels, Belgium }


\begin{abstract}
We consider a single molecule circuit embedded into solvent. The Born dielectric solvation model
is combined with  Keldysh nonequilibrium Green's functions to describe the electron transport properties of the system.
 Depending on the dielectric constant, the solvent induces
multiple nonequilibrium steady states with corresponding hysteresis in molecular current-voltage characteristics as well as negative differential resistance.
We identify the physical range of solvent and molecular parameters where the effects are present.
The position of the negative differential resistance peak can be controlled by the dielectric constant of the solvent.
\end{abstract}

\maketitle


The use of molecules  -- either singly or in small ensembles -- as the elements of electronic circuits holds
 substantial promise in the fields of  informational technology, biological and environmental nanosensors, and  energy
harvesting.\cite{Galperin22022008} For  the science of molecular
electronics to be transformed into a technology it is not only important to fabricate stable molecular junctions
but also to be able to efficiently control and manipulate  their electric properties.
In the silicon-based  microelectronic technology the gate voltage regulates the flow of electrons, but
 placing a third gate electrode has  proven to be  difficult in single molecular size devices.
 The  negative differential resistance (NDR)  also plays an important role in  semiconductor devices, because
circuits with complicated functions can be implemented with significantly fewer components with its help.
On the other hand, instead of copying the existing paradigms, such as, for example,  gate voltage or resonant tunneling diode structure for NDR,   the molecular electronics create new and unique opportunities.
The "wet" molecular electronics, where  solvent controls the electric behavior of an electronic circuit, may open a new chapter in device engineering.
Indeed, some molecular electronic devices already exploit the solvent around the molecule
to modulate  conductance through alteration of the charge state or polarizability of the molecule.\cite{Xiao05,Chen0478474,Morales05}

\begin{figure}[t!]
\begin{center}
\includegraphics[width=\columnwidth]{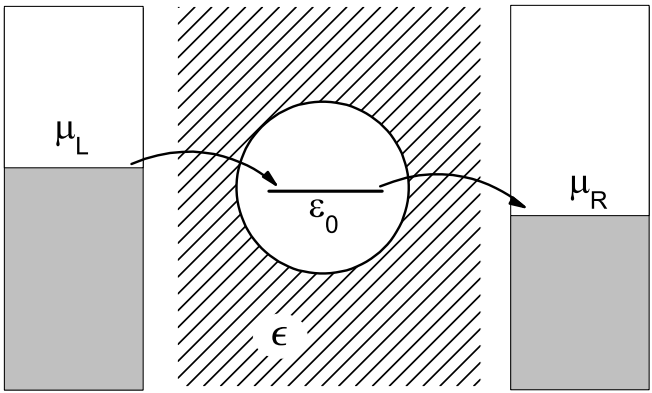}
\end{center}
\caption{Schematic illustration of the model. The molecule is attached to two metal electrodes and surrounded by solvent. The solvent is described by uniform dielectric constant
$\epsilon$.}
\label{figure1}
\end{figure}

Let us consider a "wet" molecular circuit --   a molecule attached to two macroscopic metal electrodes and embedded into solvent (Fig.~\ref{figure1}).
The total Hamiltonian is
\begin{equation}
 {H}= {H}_{L}+ {H}_{R}+ {H}_{M}+ {H}_{T} +  {H}_{MS}.
\end{equation}
The left and right electrodes contain free electrons and are described by the following Hamiltonians:
\begin{equation}
  {H}_{L}=\sum_{l \sigma} \varepsilon_{l}a_{l\sigma }^{\dagger}a_{l \sigma} , \;\;\;\;\  {H}_{R}=\sum_{r \sigma }\varepsilon_{r }a_{r \sigma }^{\dagger}a_{r \sigma}.
 \end{equation}
 Here $a^\dagger_{l\sigma/r \sigma}$   creates  an electron with spin $\sigma$ in the single-particle state $l/r$ of the left/right electrode and $a_{l\sigma/r \sigma}$ is the corresponding electron  annihilation operator.
The molecule is described by a single spin degenerate electronic level with energy $\varepsilon_0$
\begin{equation}
 {H}_M  = \varepsilon_0  \sum_{\sigma} a^\dag_\sigma a_\sigma.
\end{equation}
The operator $a^\dag_\sigma (a_\sigma)$ creates (destroys) an electron with spin $\sigma$ on the molecular level.
The tunneling coupling between the molecule and electrodes is
 \begin{align}
  {H}_{T}=  \sum_{l\sigma}t_l (a_{l \sigma}^{\dagger}a_{\sigma }+ h.c ) +  \sum_{r\sigma} t_r (a_{r \sigma}^{\dagger}a_{\sigma }+ h.c ).
\end{align}
The interaction between the molecule and the surrounding solvent, $H_{MS} $,  will be discussed below.
We use natural units in equations throughout the paper:
$\hbar = k_B = |-e| = 1$, where $-|e|$ is the electron charge.

We describe the interaction between the molecule and the solvent based on the following simple model. The molecule is considered as a conducting sphere of radius $R$ and the solvent is macroscopically uniform and characterized by dielectric constant $\epsilon$. The work needed to place charge $q_M$ on  a conducting sphere in the dielectric environment is given by the Born expression for the dielectric solvation energy \cite{nitzan-book}:
\begin{equation}
W  =  \frac{q_M  q_S}{2R} \left( 1 - \frac{1}{\epsilon} \right),
\end{equation}
where $q_S$ is the induced charge in the solvent ($q_M = - q_S$).
The model  can be easily extended to  the molecules of complex shapes (the so-called generalized Born model, which represents the molecule as a number of overlapping  spheres of different radii).\cite{generalized-born}
The (generalized)  Born model is quite simple yet is very successful in computing the electrostatic contribution to the solvation free energy.\cite{leach-book,generalized-born}
The solvent dynamics is slow in comparison with the electron tunneling time scale. For example, the dielectric relaxation of the solvent is diffusive and occurs on the  picosecond or slower  time scales since dipolar solvent molecules generally  respond to the change of  the molecule junction charging state by rotating.\cite{nitzan-book} Therefore, we can assume that the induced charge $q_S$ corresponds to the average electronic population of the molecular junction.
Then,  the dielectric solvation energy  can be directly associated with the interaction of the molecule with the surrounding solvent:\begin{equation}
 {H}_{MS} = -U(\epsilon)( {N} - \delta) (\langle N \rangle - \delta),
 \label{mf}
\end{equation}
where $U(\epsilon) =   \frac{1}{2R} \left( 1 - \frac{1}{\epsilon} \right)$ is an effective, local and solvent controlled electron-electron attraction,
and $  N=\sum_\sigma a^\dag_\sigma a_\sigma$.
The charge of the molecule due to nonequilibrium tunneling of electrons is $( {N} - \delta) $, while
$(\delta -\langle N \rangle)$ is the corresponding induced charge in the solvent.
The parameter $\delta$ is the equilibrium molecular electronic population
which depends on the position of the molecular level $\varepsilon_0$ relative to the electrode Fermi energy $\varepsilon_f $.
If $\varepsilon_0$ corresponds to the highest occupied molecular orbital  (i.e., $\varepsilon_0 < \varepsilon_f$), then, without the applied voltage bias, the molecular level is double occupied and $\delta=2$.
If $\varepsilon_0$ is the  lowest unoccupied molecular orbital (i.e. $\varepsilon_0 > \varepsilon_f$), then the molecular level is empty in equilibrium and $\delta=0$.
It is known that such model Hamiltonians generally lead to bistable solutions.\cite{PhysRevB.67.075301,galperin05}
We emphasize that the model is not only applicable to the solvated molecular junction but also to the often employed experimental setting when the junction is embedded into isolating or semiconductor molecular film. In this case the surrounding molecular film can be considered as a macroscopic dielectric environment.

The similar mean-field-type  interaction between the molecule and the solvent (Eq.~\ref{mf}) can be also obtained within the polaron model in the limit $\omega/\Gamma << 1$  (here $\omega$ is  the frequency of a characteristic vibrational mode coupled to the electrons and $\Gamma$ is the broadening of the molecular level due to coupling to the metal electrodes).\cite{kuznetsov-ndr,galperin08}
 In our case $\omega$ is related to the dielectric relaxation of the solvent, which  occurs
on  the picosecond and  slower time scales, so $\omega \sim 0.001$ eV. For molecules interacting with the metal electrodes, $\Gamma \sim 0.1-1$ eV, which makes the static, effective mean-field (i.e., the static, average polarization of the solvent) approximation  (Eq.~\ref{mf})
exactly valid for our case.

\begin{figure}[b!]
\begin{center}
\includegraphics[width=\columnwidth]{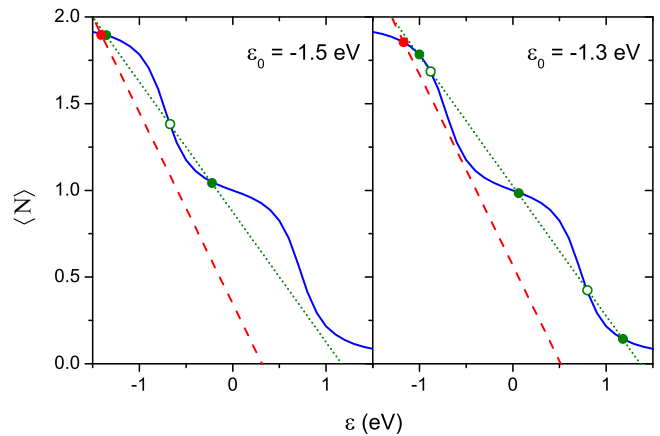}
\end{center}
\caption{Graphical solution of Eq.~\eqref{population}.  The straight lines are given by equation $\langle N \rangle   = -\frac{\varepsilon}{U(\epsilon)} + \Bigl(2 + \frac{\varepsilon_0}{U(\epsilon)}\Bigr)$.   Depending on $\epsilon$ and $\varepsilon_0$ there can  exist one, three, or five fixed  point nonequilibrium molecular populations.
Filled circles represent stable steady state populations, while open ones correspond to unstable fixed point solutions of Eq.~\eqref{population}.
Parameters: applied voltage bias $V=1.5$~eV, $T=300$~K, $\Gamma_0=0.1$~eV,  $R=10$~Bohr,  $\epsilon=50$ (dotted lines), and $\epsilon=3$ (dashed lines). }
\label{figure2}
\end{figure}

Thus, in the Born approximation,  the Hamiltonian $H_M + H_{MS}$  is exactly reduced to a  spin degenerate single-level model with a local mean-field  attractive
interaction between electrons, which can be controlled by the dielectric constant of the environment.
To describe electron transport through the system we use Keldysh nonequilibrium Green's-function formalism.\cite{keldysh65, haug-jauho}
The exact nonequilibrium molecular population $\langle N\rangle$ and electric current $J$ become:
\begin{equation}
 \langle N\rangle  = \frac{2}{\pi}\int d\omega \frac{ \Gamma_L(\omega) f_L(\omega) + \Gamma_R(\omega) f_R(\omega) }{ (\omega - \varepsilon - 2\Lambda(\omega))^2 + (\Gamma(\omega))^2}
  \label{population}
\end{equation}
\begin{equation}
  J=\frac{4}{\pi} \int d\omega \frac{\Gamma_L(\omega)\Gamma_R(\omega) (f_L(\omega)-f_R(\omega))}{(\omega-\varepsilon-2\Lambda (\omega))^2 +(\Gamma(\omega))^2}.
\end{equation}
Here $f_{L/R}(\omega)= [1+e^{(\omega - \mu_{L/R})/T}]^{-1}$ is the Fermi-Dirac distribution for electrons in the electrodes,
$ \varepsilon =   \varepsilon_0 - U(\epsilon)(\langle N \rangle - \delta)$ is the effective energy of the molecular level, and
$\Lambda=\Lambda_{L}+\Lambda_{R}$, $\Gamma=\Gamma_L+\Gamma_R$ are the real and imaginary parts of the electrode self-energy
\begin{equation}\label{SE}
  \Sigma_{L/R}= \sum_{k\in l/r}\frac{t_k^2}{\omega-\varepsilon_k +i\eta} = \Lambda_{L/R}(\omega) - i\Gamma_{L/R}(\omega).
\end{equation}
The electrodes are modeled as a semi-infinite chain of atoms, characterized by the voltage-dependent on-site energy $\mu_{L,R}=\pm V/2$ and the
inter-site hoping parameter $V_h=2.5$~eV. The expression for the electrode self-energy can be found, for example, in~\cite{peskin2010}.
The electrode bandwidth $[\mu_{L/R}-2V_h,\mu_{L/R}+2V_h]$ is half filled, so the Fermi energy coincides with
the one-site energy.
The coupling between the left/right electrode edge and the molecule is taken  to be
$\sqrt{V_h \Gamma_0}$, where $\Gamma_0=\Gamma_L(\mu_L)=\Gamma_R(\mu_R)$ is the maximal broadening of the molecular electronic level
due to the coupling to the  electrodes.
Below we focus on the case  when $\varepsilon_0$ is lower  than the electrode equilibrium  Fermi energy.
All our results also remain qualitatively valid when $\varepsilon_0 $ is above the Fermi level.

\begin{figure}[b!]
\begin{center}
\includegraphics[width=\columnwidth]{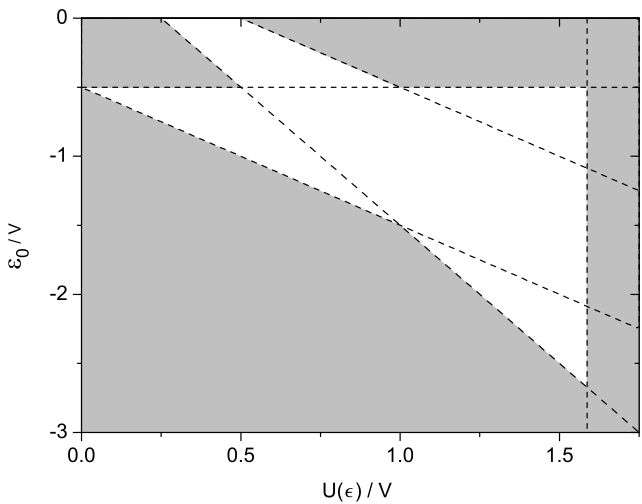}
\end{center}
\caption{The white domain corresponds to the values of $\varepsilon_0$ and $U(\epsilon)$  where the multiple steady state solutions exist ($\Gamma_0/V << 1$).
Dashed lines determine domain boundary and they are
$\varepsilon_0 = - 0.5V$,   $\varepsilon_0 = - U+ 0.5V $, $\varepsilon_0 = - U - 0.5V$,  $\varepsilon_0 = -2U+0.5V$, and $U = 1/2R$. }
\label{figure3}
\end{figure}

To compute the current, we first  should determine the nonequilibrium molecular population $\langle N\rangle$.
Since Eq.~\eqref{population}
is nonlinear, it generally has multiple solutions.
 Fig.~\ref{figure2} shows the graphical solution of this equation. As we see,  likewise for the electron transport in the polaron model \cite{galperin05}, depending on values of $U(\epsilon)$ and the molecular level energy $\varepsilon_0$, Eq.~\eqref{population} can
have one, three, or even  five solutions (the nonequilibrium fixed points).
These multiple solutions
may or may not be steady states (i.e., the stable fixed point).
Following our method described  in~\cite{stability}
we obtain the stability matrix  and analyze the real part of its spectrum to assess the asymptotic time behavior of the fixed points.
We find that
only the two outer and the middle solutions are stable in the five-solution case  (right panel in Fig.\ref{figure2}); i.e., they  correspond to physically realizable nonequilibrium steady-state populations. In the case of  three solutions (left panel in Fig.\ref{figure2}), the middle
solution is unstable and the other two fixed points are stable.
We note that our approach is immune from the criticism that the observed multiple steady states  are artifacts of the mean-field and electron self-interaction.\cite{alexandrov06}
The effect of self-interaction is physically  present in our case, since an electron in the molecule interacts with its own induced charge in the solvent.

Let us now  establish the range of key physical parameters -- dielectric constant $\epsilon$, molecular size $R$, and molecular level energy $\varepsilon_0$, which allow the existence of  multiple nonequilibrium steady states. For presentation purposes  we assume that the molecular level broadening, $\Gamma_0$, as well as the temperature are  much smaller than applied voltage $V$.
Therefore
the molecular population (Eq.~\ref{population}) (solid lines in Fig.~\ref{figure2}) can be approximated by a step like function of energy $\varepsilon$.
Then, we can readily determine analytically the conditions on $\varepsilon_0$ and $U(\epsilon)$ when Eq.~\eqref{population} has only one solution. In Fig.~\ref{figure3} we show the domain where multiple steady states exist for the case $\varepsilon_0<0$. The case $\varepsilon_0>0$ can be considered in the same way, and the resulting multistability domain is a mirror reflection of that in Fig.~\ref{figure3} across the abscissa axis.

\begin{figure}[t!]
\begin{center}
\includegraphics[width=\columnwidth]{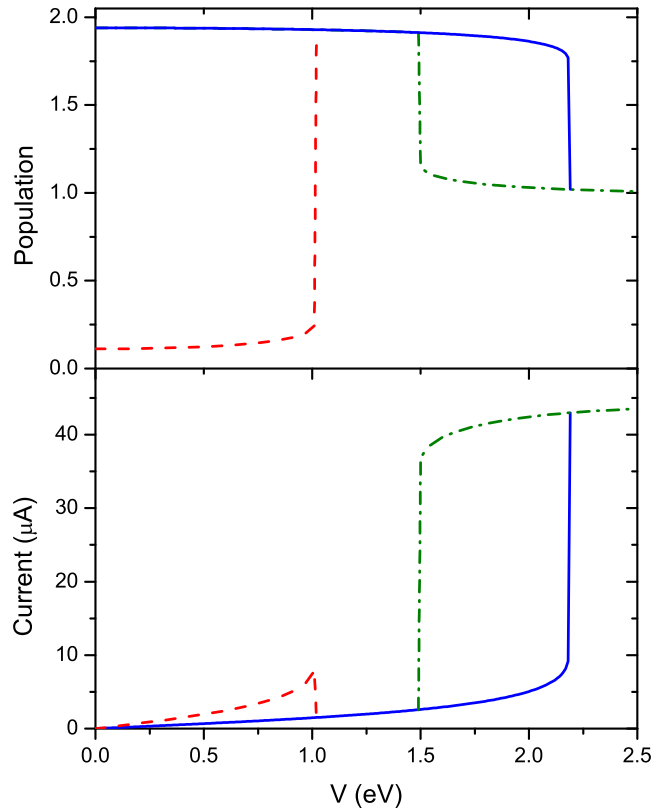}
\end{center}
\caption{Population-voltage and current-voltages characteristics. Parameters are $T=300$~K, $\varepsilon_0=-1.6$~eV, $R=10$~Bohr, $\epsilon=50$, $\Gamma_0=0.1$~eV.
Three curves correspond to  three possible roots of Eq.\eqref{population}: solid line - upper root, dashed line - lower root, and dash-dotted line --
middle root.}
\label{figure4}
\end{figure}

In Fig.~\ref{figure4}, we show the behavior of the level population and the current as a function of applied voltage.
Due to the presence of multiple steady states, both the population and the electron current demonstrate a hysteresis behavior.
The width of the hysteresis loop is proportional to $U(\epsilon)$ and, therefore,  it can be controlled by the dielectric constant. It should be emphasized that  the solvent-induced  hysteresis loop can be observed at moderate applied voltages where the molecular device
is still mechanically stable. Moreover,
the nonlinearity in the molecule-solvent interaction leads to  NDR features in the  current - voltage characteristic (the drop in the current represented by the dashed line at around 1 eV of applied voltage in  Fig.~\ref{figure4}).
The NDR  appears when one of the electrode chemical potentials crosses the position of the molecular level. Then, due to the subsequent shift in the level energy caused by the electronic population change, the level moves
away from the current-carrying  window between  the chemical potentials. In the case of $\varepsilon_0<-0.5U(\epsilon)$ shown in Fig.~\ref{figure4}  the NDR takes place when we begin with the empty level.
When $\varepsilon_0$ lays above $-0.5U(\epsilon)$ ($\varepsilon_0<0$) the NDR also takes place, but in this case we need to start from the initially fully occupied level.

\begin{figure}[t!]
\begin{center}
\includegraphics[width=\columnwidth]{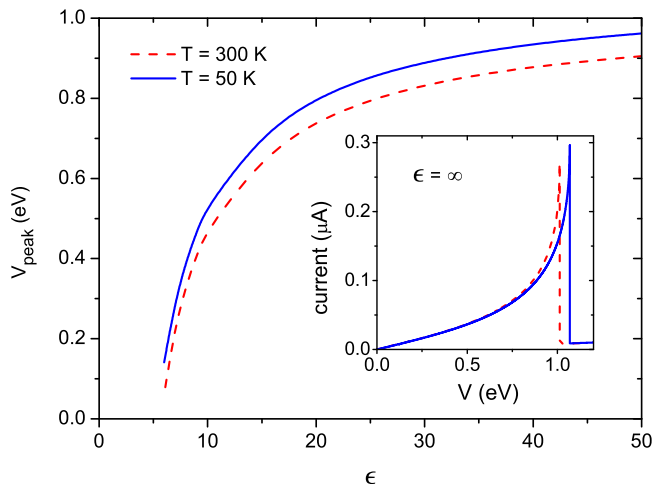}
\end{center}
\caption{
The NDR peak voltage value as a function of the dielectric constant for two different temperatures. Parameters are
 $\varepsilon_0=-2.0$~eV, $R=10$~Bohr, $\Gamma_0=0.01$~eV. Inset:  NDR in current-voltage characteristic for $\epsilon=\infty$.}
\label{figure5}
\end{figure}

The NDR in the  "wet" molecular circuit turns out to be  sensitive to the dielectric constant of the environment. Figure~\ref{figure5} shows the
dependence of the NDR peak  position on the dielectric constant of the solvent.
The increase of the solvent polarity shifts the peak toward the higher voltages.
This effect is very robust. It does not require an artificial tuning of the model parameters  and holds at very large ranges of temperatures.
The temperature dependence of the NDR peak (inset in Fig.\ref{figure5}) is consistent with experimental observations,\cite{Chen99a,Chen02} and in contrast to the polaron model explanation of  NDR \cite{galperin05} does not require unphysical values for the parameters.

We would like to comment here on the importance of the time scales. Depending on the relative time scales of measurements and transitions between stable fixed points, the multistability can result in merely noise associated with the jumps between steady states or it can lead to hysteresis and NDR~\cite{emanuel2006}. To be experimentally resolved the transition rate  between multiple steady states should be smaller than the typical observation time. In our case the transition between steady states is determined by the very slow diffusive reorganization of the solvent, which opens a possibility for experimental realization of the proposed effects.

In conclusion, we have presented a  theoretical model to describe the environmental control of the electron-transport properties of "wet" molecular junctions.
The interaction between the molecule and solvent leads to effective attraction between
electrons which is governed by the dielectric constant of the surrounding solvent.
The natural separation of electronic and solvent time scales
makes the  mean-field consideration exact for our model.
We used Keldysh nonequilibrium Green's functions to obtain a nonlinear equation for molecular population and electric current.
Depending on the dielectric constant, the inherent nonlinearity of molecule-solvent interactions induces
multiple nonequilibrium steady states with corresponding hysteresis in molecular I-V characteristics as well as NDR.
We identify the physical range of solvent and molecular parameters which allows the appearance of multiple steady states.
The temperature effects on the NDR peak are in qualitative agreement with the available experimental data.
We demonstrated that the dielectric constant of the solvent can be used as a control parameter which regulates the position of the NDR peak.

\begin{acknowledgments}
 This work has been supported by the Francqui Foundation, Belgian Federal Government, under the Inter-University Attraction Pole project NOSY  and
 Programme d'Actions de Recherche Concert\'ee de la Communaut\'e Fran\c~caise (Belgium), under project "Theoretical and experimental approaches to surface reactions".

\end{acknowledgments}


\end{document}